\documentclass[prl,aps,graphics,a4paper,10pt,twocolumn,superscriptaddress]{revtex4-1} 

\usepackage[margin=25mm]{geometry}

\usepackage{graphicx}
\usepackage{mathrsfs}
\usepackage{xfrac}
\usepackage{amsmath} 
\usepackage{amsfonts} 
\usepackage{braket}
\usepackage{color}
\usepackage{placeins}
\usepackage{hyperref}
\usepackage{tabu}
\usepackage{colortbl}
\usepackage[dvipsnames]{xcolor}
\usepackage{soul}
\usepackage{algorithmic}
\usepackage{ulem}
\usepackage{scrextend}

\newcommand{\robnote}[1]{}

\newcommand{\be}{ \begin{equation} \begin{split} }
\newcommand{\ee}{ \end{split} \end{equation}  }

\newcommand{\nm}[1]{M_j}

\newcommand{\Ab}{A}
\newcommand{\Em}{E}

\newcommand{\gd}{\Gamma_{\downarrow}}

\newcommand{\proj}[1]{\mathcal{P}_{#1}}
\newcommand{\A}[1]{B_{#1}}

\newcommand{\dvecf}[1]{\dot f_{#1}}

\newcommand{\Imperial}{
Physics Department, Blackett Laboratory, Imperial College London, \\Prince Consort Road, SW7 2AZ, United Kingdom}
\newcommand{\ImperialCQD}{Centre for Doctoral Training in Controlled Quantum Dynamics, Imperial College London, \\Prince Consort Road, SW7 2AZ, United Kingdom}

\begin{document}

\title{Non-critical slowing down of photonic condensation}

\author{Benjamin T. Walker}\affiliation{\Imperial}\affiliation{\ImperialCQD}
\author{Henry J. Hesten}\affiliation{\Imperial}\affiliation{\ImperialCQD}
\author{Himadri S. Dhar}\affiliation{\Imperial}
\author{Robert A. Nyman}\affiliation{\Imperial}
\author{Florian Mintert}\affiliation{\Imperial}


\begin{abstract}
We investigate the response of a photonic gas interacting with a reservoir of pumped dye-molecules to quenches in the pump power.
In addition to the expected dramatic critical slowing down of the equilibration time around phase transitions we find extremely slow equilibration even far away from phase transitions.
This non-critical slowing down can be accounted for quantitatively by fierce competition among cavity modes for access to the molecular environment,
and we provide a quantitative explanation for this non-critical slowing down.
\end{abstract}

\maketitle

The time scales of evolution of simple dynamical systems are typically directly related to system parameters.
By contrast, systems with many degrees of freedom can exhibit emergent behavior, with dynamics on time-scales that have no clear origin in the microscopic equations of motion.
A prominent example is the relaxation towards a steady state which is known to slow down when a system is close to a critical point ~\cite{HohenbergHalperin77,suzuki1981phase}.
This critical slowing down is present in the statistical mechanics of systems from atomic quantum gases~\cite{labouvie2016} or magnetic metamaterials~\cite{anghinolfi2015} to entire ecosystems or human societies~\cite{dakos2010}.

The slowing down of system dynamics goes hand-in-hand with an increase in the amplitude of the fluctuations in both classical~\cite{kawasaki1970}
and quantum statistical systems~\cite{clark2016}.
Close to criticality, fluctuations and dynamics become linked, and the system behavior can be characterized by critical exponents~\cite{HohenbergHalperin77,navon2015}.
In this respect, critical slowing down is a very different phenomenon from other delayed-equilibrium phenomena such as prethermalization of isolated quantum gases~\cite{gring2012} or light~\cite{santic2018}, and Anderson localization~\cite{Billy2008}. Non-critical slowing down is typically related to the integrability of the dynamical variables or to the kinetic impossibility of exploring the full state space available, and therefore not directly linked to fluctuations.

In this manuscript we investigate the slowing down of equilibration in a photonic gas in a pumped, dye-filled optical microcavity.
Far from critical pump parameters, the response of the photon populations to abrupt changes in pumping has been shown experimentally to occur on timescales typical for the absorption of a cavity photon by a dye-molecule~\cite{schmitt2015thermalization,hakala2018bose}. The fluctuations of photon numbers~\cite{schmitt2014observation} and the phase of a Bose-Einstein condensate~\cite{walker2017driven, schmitt2016} have been observed to occur on similar timescales. Here, we find both critical and non-critical slowing down. Unusually, the non-critical slowing down does not appear to be related to an impossibility of exploring the state space, but rather to a detailed balance of excitations being exchanged between photon modes and partially-overlapping subsets of the dye molecules.
Furthermore, it is connected to an inversion of an important susceptibility of the system (derivative of photon number in a specific mode with respect to pump power), implying a very unusual relationship between equilibration and fluctuation.

The system dynamics is described well in terms of rate equations for the occupations $n_i$ of cavity modes and a vector $f$ characterizing the inhomogeneous fractional excitation of all dye-molecules~\cite{keeling2016}.
A suitable construction of collective modes of molecular excitations~\cite{slowPRA} enables the reduction of the molecular environment to its most relevant degrees of freedom.
To this end, the vector $f$ is decomposed into a hierarchy of several components $f_j=\proj{j}f$ with suitably defined projectors $\proj{j}$.
By construction 
the cavity dynamics depends only on the level-$0$ component $f_0$,
and the level-$(j+1)$ component $f_{j+1}$ affects the cavity dynamics only indirectly via its influence on the level-$j$ component $f_j$.
A truncation after $2$ or $3$ levels provides an excellent description of the system dynamics with substantial gain in numerical efficiency~\cite{slowPRA}.

The level-$0$ component $f_0$ can be expanded into a set of vectors ${\bf e}_i$ such that the equation of motion~\cite{keeling2016,slowPRA}
\begin{equation}
\dot{n}_i =[n_i (\Em_i+\Ab_i)+\Em_i] c_iv_i-\gamma_i n_i\ ,
\label{eq:ndot}
\end{equation}
for each occupation number $n_i$ of a cavity mode depends only on the single component $v_i={\bf e}_i^T f_0$ of the molecular environment.
The corresponding coupling constant $c_i$ reads $c_i=\sum_jg_{ij}\nm{j}[{\bf e}_i]_j$, where $g_{ij}$ is the coupling constant between cavity mode $i$ and a dye molecule at position $j$;
$\nm{j}$ 
is the number of molecules in the small volume element around this position, and $[{\bf e}_i]_j$ is the element $j$ of the vector ${\bf e}_i$.
$\Ab_i$ and $\Em_i$ are the rates of absorption and emission of a dye molecule,
and $\gamma_i=\Ab_i \sum_j g_{ij}\nm{j}+\kappa$ is the decay constant, including the cavity decay rate $\kappa$.

The equations of motion for the excited state fractions of molecules $f_j$ read
\begin{equation}
\begin{split}
\dvecf{j}&=\sum_{k}\sum_{i} n_i\proj{j}\A{i}\proj{k}f_k-\A{0}f_j  - \proj{j}x\ ,
\label{eq:dyn}
\end{split}
\end{equation}
in terms of the diagonal matrices $\A{i}$ with diagonal elements
$[\A{i}]_{pp}=-(\Em_i+\Ab_i)g_{ip}$,
the matrix
$\A{0}=\sum_{i}\A{i}\Em_i/(\Em_i+\Ab_i)+(\gd+P)\mathbf{1}$,
the vector
$x$ with elements
$[x]_j=P+\sum_{i} g_{ij} \Ab_i n_i$,
the pump rate $P$, and the decay constant $\gd$ for non-radiative decay or emission into free space~\cite{keeling2016,slowPRA}.

This approach allows us to investigate the equilibration of the light in the cavity after quenches in pump power.
Strictly speaking, a steady state is reached only asymptotically,
but in practice, one can accept small deviations and thus define a finite time to reach stationarity.
We deem stationarity to be reached if the difference between populations of current state and exact steady state reaches a specified fraction of the exact steady state population for each mode of the cavity.
The specification of this fraction is largely arbitrary; we chose a value of $10^{-6}$, and a different choice would result in an overall change of time-scale.

We consider a two-dimensional cavity with parabolic mirrors and harmonic oscillator eigenmodes labeled with the double index $i=[m_x,m_y]$.
{We take into account the lowest 5 energy levels, corresponding to 15 photonic cavity modes. In the Supplementary Material (SM)~\cite{supmat2} we show that the numerical results for these parameters are unchanged for larger numbers of modes with a fixed number of molecules}. 
Expressing all system parameters in units of cavity decay constant $\kappa$ and the harmonic oscillator length $l_{ho}$,
we use a molecular density of $10^{13}/l_{ho}^{2}$, $\nm{j}=10^{12}$
molecules in each group and a molecular decay rate $\gd =\kappa/4$; {the absorption and emission rates are $A_{[m_x,m_y]}= 10^{-12}\kappa\ [3.8,\ 9.2,\ 23.0,\ 55.4,\ 124.9]_{m_x+m_y}$, and $E_{[m_x,m_y]} = 10^{-10}\kappa\ [5.6,\ 6.8,\ 8.2,\ 9.3,\ 10.0]_{m_x+m_y}$, respectively, matching that of rhodamine~$6$G.}

%

\begin{figure}[t]
\centering
\includegraphics[width=0.46\textwidth]{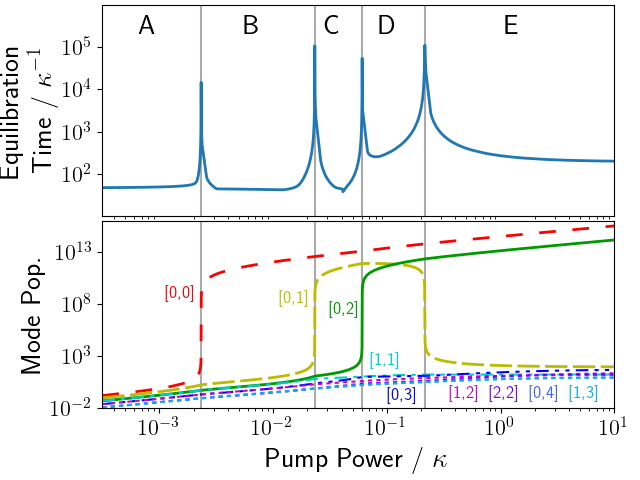}
\caption{(TOP) The time taken for the system to equilibrate after a quench in pump power by $1\%$, as function of the pump power after the quench.
(BOTTOM) The steady populations $n_i$ of cavity modes i ranging from $[0,0]$ to $[1,2]$ and $[0,3]$. The mode populations feature sharp increases and drops under increase of pump power, and the equilibration times show clear peaks around those phase transitions.
In addition to this, the equilibration is also strongly slowed down in the intervals labeled `$D$' and `$E$' far away from any known phase transition.}
\label{fig:off_diag}
\end{figure}

In order to analyze the equilibration time, we start with the system in its stationary state at a given pump power and then increase the pump power with a quench of $1\%$.
Fig.~\ref{fig:off_diag}~(TOP) shows the time taken to reach the steady state after this quench as function of post-quench pump power;
the bottom panel shows the corresponding steady state population in each mode at the final pump power.
One can see clear peaks in equilibration time at pump powers at which cavity modes condense or decondense.
The intervals below and between the phase transitions seen in Fig.~\ref{fig:off_diag}~(TOP) are labeled by letters `A' to `E'.
In the intervals `A' to `C', the equilibration time is about a factor of $10$ larger than the cavity decay time $1/\kappa$.
In interval `D', however, the equilibration time does not reduce to the values found in `A' to `C', and interval `E' features a broad plateau of the equilibration time, {more than an order} of magnitude larger than the base value around $10/\kappa$. We explicitly verify that this plateau extends {to pump rates reaching $10^3 \kappa$}, and that for parameters where a further condensation peak is observed, the {non-critical slowing is not dependent on the tails of a subsequent condensation peak} (see the SM in Ref.~\cite{supmat2}).

Critical slowing down around phase transitions is well known \cite{labouvie2016,anghinolfi2015,dakos2010},
and we {find that all the four phase transitions are characterized} by the same critical exponent of $1$, {\it i.e.} a divergence of the equilibration time $\propto |P-P_{c}|^{-1}$ as the pump power $P$ reaches its critical value $P_{c}$.
In contrast to this well known critical slowing down, however,
the increase of equilibration time in `D' and `E' is not associated with any phase transition.
To the best of our knowledge, such {\it non-critical} slowing has never been observed, and quite strikingly the time-scale {more than $10^{2}/\kappa$} does not match any of the natural time-scales of the system.

Both the fast and the slow equilibration times can be explained in terms of Eq.~\eqref{eq:ndot}.
Photons are being lost from the system with rate $\kappa$, and since the exchange of photons between cavity and environment is faster than this loss process, the system relaxes to the new steady state with a decay constant close to $\kappa$ after the change in pump power.
Assuming an exponential decay, the relative deviation $\delta n_i$ from the steady state has decayed to a value of $d$ after the time $t_e$ satisfying $\delta n_i\exp(-\kappa t_e)=d$.
With the value of $\delta n_i=2\%$ that we find for a quench by $1\%$ sufficiently far away from the phase transitions, and the threshold $d=10^{-6}$, one would thus expect an equilibration time of $t_e=-\ln(5\times10^{-5})/\kappa\simeq 10/\kappa$ which matches the observed values very well.

The slowing down of equilibration time can be attributed to the fact that the molecular excitation $v_i$ remains close to its critical value despite the quench in pump power.
If this is the case, it is convenient to re-express Eq.~\eqref{eq:ndot} as
\begin{equation}
\dot{n}_i =-(E_i+A_i) (v_i^{c}-v_i)n_i+E_iv_i\ ,
\label{eq:ndotfc}
\end{equation}
in terms of the critical excitation $v_i^c=\gamma_i/(E_i+A_i)$ for which the stationary state solution of Eq.~\eqref{eq:ndot} diverges.
If $v_i$ remains constant, one can understand $\eta_i=(E_i+A_i)(v_i^{c}-v_i)$ as the effective decay rate,
and whenever the dye excitation $v_i$ approaches its critical excitation $v_i^{(c)}$, the effective rate $\eta_i$ becomes minute and the dynamics of mode $i$ slows down.

\begin{figure}[t]
\includegraphics[width=0.45\textwidth]{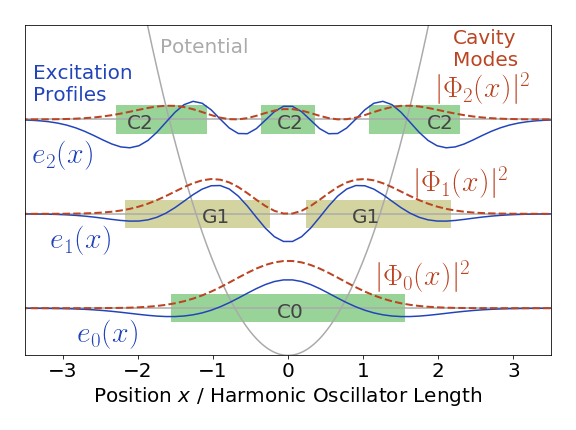}
\caption{
	Moduli squares $|\Phi_i(x)|^2$ for $i=0,1,2$ of the three lowest eigenfunctions of a quantum harmonic oscillator are depicted in red.
	The corresponding harmonic potential and lines indicating the eigen-energies are depicted in grey.
	The excitation profile functions ${\bf e}_i(x)$ corresponding to each of these cavity modes are depicted in blue.
	Green bars (`C0'/`C2') approximate one-dimensional cuts through the areas in which modes $[0,0]$ and $[0,2]$ compete for excitations and clamp molecules when condensed. Yellow bars (`G1') approximate the locations of molecules absorbing/emitting photons from/into mode $[0,1]$. Mode $[0,1]$ experiences competition through the whole of G1 with either $[0,0]$ or $[0,2]$.}
\label{fig:molecularmodes}
\end{figure}

Consistent with critical slowing down,
the condition $v_i\lesssim v_i^{(c)}$ is typically satisfied close to a phase transition, even after a quench in pump power.
Far away from a phase transition,
any quench in pump power causes $v_i$ to differ substantially from its critical value so that regular equilibration on the timescale $\kappa$ applies, but
as one can see in Fig.~\ref{fig:off_diag} this does not hold in the intervals `D' and `E'.

In order to understand why the excitations $v_{[0,1]}$ in the molecular environment that are accessible to the cavity mode $[0,1]$, hardly react to quenches in pump power in the intervals `$D$' and `$E$', but not in any of the other intervals, one has to inspect the effect of clamping caused by the condensed modes.
Given the macroscopic occupation of condensed modes, the coupling between those modes and the molecular environment is extremely strong.
If this coupling is so strong that the interaction between a molecule and the external pumping becomes negligible, this molecule will not increase its excitation as pumping is increased, and it is considered to be clamped to the condensed mode \cite{keeling2016,PhysRevLett.120.040601}.

The mode-vectors ${\bf e}_i$ that describe the spatial excitation profile to which the modes $i$ couple, share similarities with the spatial profile of the cavity modes, as one can see in Fig.~\ref{fig:molecularmodes}, where the mode profiles for the three lowest eigenstates of a harmonic oscillator and the corresponding excitation profiles ${\bf e}_i$ are depicted.
Since those profiles coincide with one-dimensional cuts through the excitation profiles ${\bf e}_{[0,i]}$, Fig.~\ref{fig:molecularmodes} provides a physical picture of
how clamping causes slow equilibration in `$D$' and `$E$', but not in the other intervals depicted in Fig.~\ref{fig:off_diag}.
Crucially, unlike the photonic cavity modes, the excitation profiles are {\it not} mutually orthogonal, and their dynamics are coupled as described by Eq.~\eqref{eq:dyn}.
This means that a cavity mode can clamp other excitation profiles as well as its own.
As one can see, the cavity mode profile of the mode $[0,1]$ has two maxima around which it couples strongly to the molecular environment (opaque regions $G1$ in Fig.~\ref{fig:molecularmodes}).
The mode $[0,0]$ couples strongly in the region $C0$ between those maxima, and the mode $[0,2]$ couples strongly between and outside those two maxima as indicated by $C2$.
There are thus regions in which the competition between mode $[0,1]$ and the modes $[0,0]$ and $[0,2]$ for access to the excitations of the dye molecules is particularly strong.
In the intervals `$B$' and `$C$', where mode $[0,0]$ is condensed, the molecules in $C0$ --- in particular in the overlap with $G1$  ---
can become clamped, but the molecules outside the maxima can still change their excitation sufficiently well in response to a quench in pump, to allow for equilibration on the regular time-scale.
In the intervals `$D$' and `$E$', mode $[0,2]$ can also contribute to the clamping, so that the molecules on both sides of the relevant domain --- including the overlap between $G1$ and $C2$ --- are being clamped.
This then results in the stabilization of $v_{[0,1]}$ close to $v_{[0,1]}^c$ and the corresponding slow equilibration, even though the driving pump has good overlap with all molecular excitation profiles.

\begin{figure}[tb]
\centering
\includegraphics[width=0.46\textwidth]{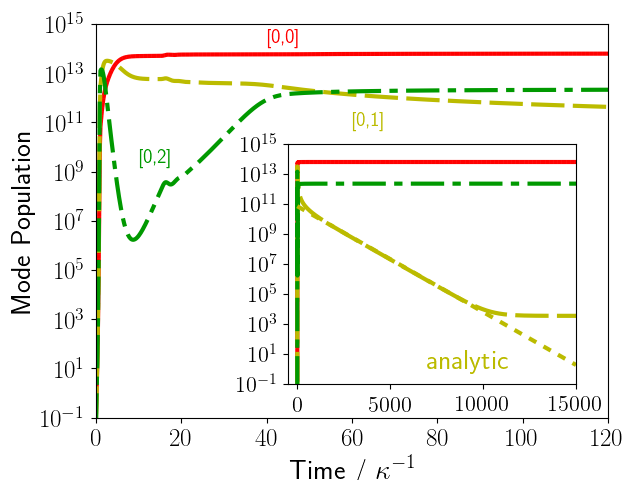}
\caption{Mode populations as a function of time after a quench in pump power from $3.16\times10^{-4}\kappa$ {to $2.5\times10^{-1}\kappa$.}
Modes $[0,0]$ and $[0,2]$ reach their new steady state on a time-scale of $10/\kappa$, but equilibration of mode $[0,1]$ is about three orders of magnitude slower.
The dotted line depicts the analytic prediction of Eq.~\eqref{eq:ndotfc}, that matches the simulated data very well over a range of $7$ orders of magnitude.
In this case, the equilibration time is determined by the dynamics of mode $[0,1]$.}
		\label{fig:steps}
\end{figure}

In order to substantiate that Eq.~\eqref{eq:ndotfc} indeed describes well the slow equilibration,
stronger quenches than $1\%$ are more informative, since these will result in pronounced dynamics of the cavity excitation which allows for stringent comparison with the analytic prediction.
Fig.~\ref{fig:steps} shows in detail the dynamics of a slow equilibration process resulting from an increase in pump power by three orders of magnitude.
One can see that the population $n_{[0,0]}$ of the ground-state mode grows monotonically as result of the quench and that it reaches its new steady state on the time-scale $10/\kappa$.
The population $n_{[0,2]}$ also reaches its new steady state quickly, though not monotonically.
The population $n_{[0,1]}$ of the first excited mode rapidly grows to a {value $14$ orders} of magnitude larger than its stationary value, but then reaches its new steady state on a substantially longer time-scale.
In the time-window between {$10/\kappa$ and $10^3/\kappa$} the decay is approximately algebraic $\propto t^{-\frac{3}{2}}$.
The inset depicts dynamics on a time-scale two orders of magnitude larger than the main figure, and confirms the subsequent
exponential decay predicted by Eq.~\eqref{eq:ndotfc} with $v_{[0,1]}$ taken from the stationary solution.
The nearly perfect agreement with the simulated data over $7$ orders of magnitude strongly supports the explanation of slow equilibration resulting from the close-to-critical value of $v_{[0,1]}$.

\begin{figure}[h]
\includegraphics[width=0.48\textwidth]{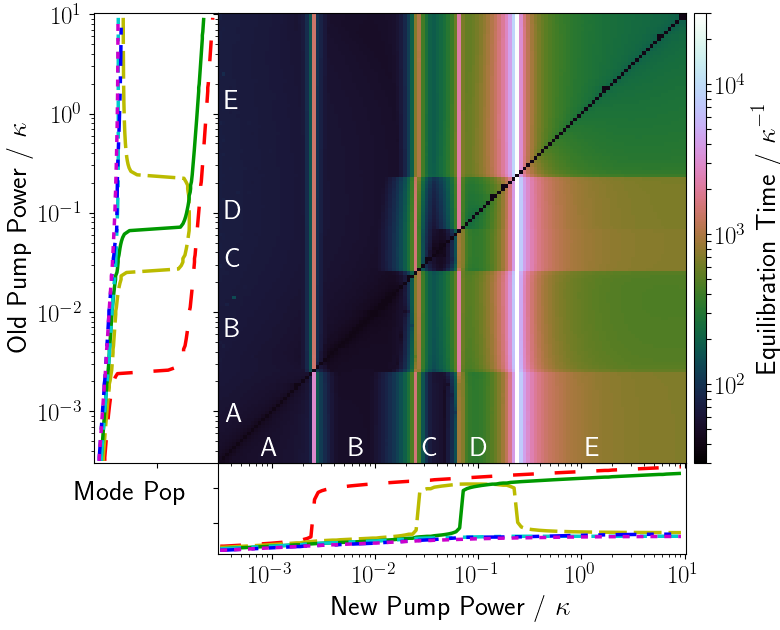}
\caption{The time taken to reach steady state after the pump power is changed from the value indicated at the left to the value depicted at the bottom of the figure.
The axes also depict the stationary state populations for the lowest three cavity modes as also shown in Fig.~\ref{fig:off_diag} and the division into intervals `$A$' to `$E$'. One can see a clear enhancement of equilibration time close to the phase transitions, {but also in the interval `$D$' and `$E$' of post-quench pump power.}} 
\label{fig:end_time}
\end{figure}

The situation depicted in Fig.~\ref{fig:steps} is not specific to the chosen values of the pump power, but rather generic, as one can see in Fig.~\ref{fig:end_time},
where the equilibration time is shown as a function of pump power before and after a quench.
The vertical lines separating the intervals `$A$' to `$E$' 
show that the equilibration time around phase transitions is slow, independent of the initial state.
Also equilibration within the intervals `$A$', `$B$', `$C$' is fast for any initial pump power.
Equilibration in the intervals `$D$' and `$E$' is always slow, and it tends to be slower in `$E$', than in `$D$' consistent with the reaction to the small quenches depicted in Fig.~\ref{fig:off_diag}.
Quite surprisingly, however, the equilibration time for a post-quench state in `$D$' or `$E$' does depend on the initial conditions to some extent.
In particular, for initial conditions in `$B$', equilibration is
less slow in `$E$', and particularly slow in `$D$'.
This can be attributed to the fact that, after a quench from initial conditions in `$A$' mode $[0,1]$ gains macroscopic occupation of approximately $10^{13}$ before being clamped by modes $[0,0]$ and $[0,2]$, as shown in Fig.~\ref{fig:steps}.
After a quench from phase `$B$', however, mode $[0,0]$ is already macroscopically occupied and immediately clamps mode $[0,1]$, which only attains a population of $10^7$ before being clamped by modes $[0,0]$ and $[0,2]$. This reduction in the population of mode $[0,1]$ reduces the time taken to decay to the steady state in `$E$', but increases the time taken to reach the macroscopic occupation of region `$D$'. 

This effect can also be exploited in order to arrive at steady states faster than with a simple quench. Starting with a pump power corresponding to interval `$A$', for example, and quenching to interval `$E$' will require an equilibration time about a factor of two longer than with a quench to interval `$B$' followed by a quench to interval `$E$' after a short delay.
This effect is just a first signature of the vast potential that temporally modulated pumping has for the control of non-equilibrium phases of light.
Varying pump powers without waiting for full equilibration will give access to the interplay of dynamics on fundamentally different time-scales that can, for example, be used to let undesired features decay while desired features are protected by slow decay times.
Together with the ability to change the effective interactions between different cavity modes in terms of suitably shaped cavity modes \cite{flatten2016} this opens entirely new avenues towards the creation of tailored states of light with abundant applications such as quantum simulations or precision sensing.
These ideas are by no means limited to bright sources of light, as considered here, but it applies equally well to systems with micro-fabricated cavities that support condensation of a few tens of photons \cite{2017arXiv170706789D} or even below ten photons \cite{walker2017driven}.
In such systems, suitably chosen temporal profiles of pumping can also be used to explore quantum states with variable inter-mode correlations and coherence properties and the suppressed interaction with the molecular environment identified here can protect such non-classical states against decoherence and decay.

We are indebted to Andre Eckardt, Alex Leymann, and Rupert Oulton for stimulating discussions.
Financial support through UK-EPSRC in terms of the Grants No. EP/ 312 J017027/1, No. EP/S000755/1 and the Centre for Doctoral Training Controlled Quantum Dynamics No. EP/L016524/1, and through the European Union's Horizon 2020 research and innovation programme under grant agreement No. 820392 (PhoQuS), is gratefully acknowledged.

\appendix*
\section{SUPPLEMENTARY MATERIAL}

In this Supplementary Material, we {provide additional analysis related to the phenomenon of non-critical slowing in photon condensates presented in the Main Text. We start in Sec.~\ref{sec1} with a
discussion on the effect of truncating our numerical simulations in the Main Text to low-energy cavity modes. We also provide evidence of the persistence of the slowing phenomena for pump powers as high as $10^3/\kappa$. In Sec.~\ref{sec2}, we show that the interesting} slowing down identified as {\it non-critical} in the Main text cannot be attributed to a long tail of the slow dynamics due to an additional phase transition resulting in {\it critical} slowing down.
In Sec.~\ref{sec3} we provide some technical details on the definition of equilibration time in our work. {We finally conclude with a short analysis on the identification of critical exponents in Sec.~\ref{sec4}.}

\section{I. Truncation of the number of modes in the cavity}
\label{sec1}

The present numerical simulations require a restriction to a small number of cavity modes which can result in truncation errors.
In general, the inclusion of more cavity modes can lead to changes in the cavity dynamics, because of the competition over the excitations in the molecular reservoirs that {\it all} cavity modes participate in.
As we will show here, {for a fixed set of system parameters, the $15$ cavity modes and a molecular environment with components up to level-2, as taken in the numerical simulations to obtain the slowing phenomenon,} captures the dynamics of all quantities of interest very well. {We note that this includes up to 3 orders (level-0, level-1 and level-2) of hierarachy in the approximation of the reduced molecular environment, as discussed in the Main Text.} 
{We compare these results with those obtained for simulations that include $10$ and $21$ modes and find that the enhanced competition in higher-energy modes does not lead to any qualitative or quantitative changes in the behavior of the system.}

\begin{figure}[h]
\includegraphics[width=0.48\textwidth]{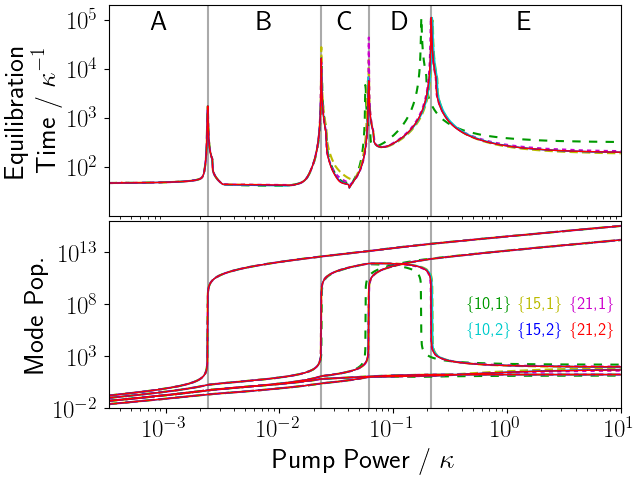}
\caption{{Simulation showing the slowing phenomenon (top) and the population (below) of the condensed and decondensed modes when simulations include $10$, $15$ and $21$ modes and molecular environment truncated up to level-1 and level-2 hierearchies. The plots here are for the same set of system parameters as those used in Fig.~1 of the Main Text, but for larger intervals in pump power. The labels $\{m,i\}$ refer to the use of $m$ cavity modes and level-$i$ hierarchy of molecular environment in the simulations.}} 
\label{10_21}
\end{figure}
\begin{figure}[h!]
\includegraphics[width=0.48\textwidth]{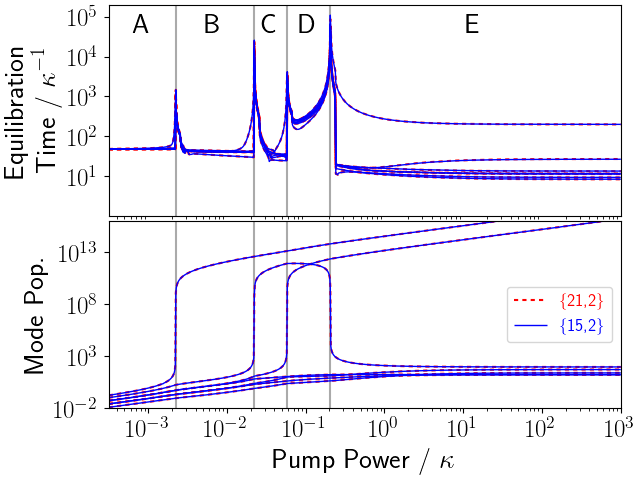}
\caption{{Equilibration time for individual cavity modes and mode populations for simulations performed using $15$ and $21$ cavity modes, with a level-2 truncated molecular environment, for pump powers as high as $10^3/\kappa$.}}
\label{15_21}
\end{figure}

{Fig.\ref{10_21}, depicts the net equilibration times together with the corresponding mode populations for simulations with $10$, $15$ and $21$ cavity modes, with the same system parameters considered in Fig.1 of the Main Text. The cut-off wavelength considered here (and in the Main Text) is $580.25$ nm and the plots are shown for both level-1 and level-2 approximations for the molecular environment. 
The simulations agree very well in the prediction for all the condensation and decondensation events taking place. For $10$ modes and level-1 molecular approximation, the observed critical pump powers are slightly different compared to the other simulations. However, excellent convergence is observed for $15$ modes and more for both the critical points and the equilibration times. In particular, the slow equilibration in interval `E' is not affected at all by further inclusion of higher cavity modes. Very similar results (not shown) are also be obtained for $28$, $36$ and $45$ modes.}
{We note that the height of the peaks in the equilibration time extremely close to the critical pump powers are dependent on how finely the points are chosen, i.e., how close one is to the critical point. This proximity can change with slight numerical changes that arise from change in system parameters such as spatial resolution of molecules or number of modes. For a sufficiently fine set of pump powers, the height in equilibration peaks at critical points will converge, but this can be numerically challenging. However, this is not relevant to explaining the non-critical equilibration times and the important slowing away from criticality, where excellent convergence is already observed.}

{Fig.\ref{15_21} shows the results obtained for simulations with $15$ and $21$ cavity modes, with level-2 approximation for the molecular environment, and now for pump powers as high as $10^3/\kappa$.
As one can see, the equilibration time and slowing behavior for all individual modes  agree very well. Moreover, for the same set of system parameters, the slow equilibration in interval `E' is persistent even for higher pump powers regardless of the number of modes in the simulation. Therefore, as justified by the analysis here, all the results in the Main Text were obtained using simulations with 15 cavity modes and molecular environment containing  up to level-2 components.}

\section{II. Non-critical slowing and further condensation peaks}
\label{sec2}

The non-critical slowing is most dramatic for dye detunings, and consequently absorption and emission rates, deep into the decondensed regime.
In this regime however there are no {additional condensation/decondensation events taking place. Therefore, no alternative explanation is possible for the observed slowing down behavior away from the critical pump powers different to the one given in the Main Text.}

\begin{figure}[t]
\includegraphics[width=0.48\textwidth]{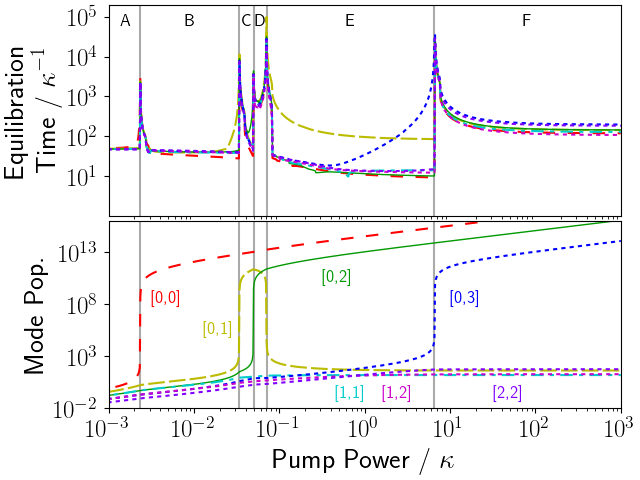}
\caption{{Equilibration time for individual cavity modes and mode populations as a function of pump power up to $10^3/\kappa$. The 
simulations here correspond to $15$ cavity modes and truncation after the level-$2$ components of the molecular environment. The label $[x,y]$ here refers to cavity modes, where $x$ and $y$ are the low-lying energy levels of the 2D harmonic oscillator, similar to those shown in Fig.~1 of the Main Text.}}
\label{3_15_v2}
\end{figure}

{In order to confirm our explanation of non-critical slowing down, we consider a different cutoff wavelength,  $579.50$~nm, with $15$ cavity modes and keeping up to level-2 components of the molecular environment in our simulation. Fig.~\ref{3_15_v2} shows equilibration times for individual modes and mode populations.}
{In addition to the slowing behavior observed so far,} there is an additional phase transition with the condensation of mode $[0,3]$ resulting in an additional instance of critical slowing down above the decondensation threshold of mode $[0,1]$.
As one can see, all modes, including mode $[0,1]$ feature critical slowing down around this phase transition, but the plateau of slow dynamics identified as non-critical slowing down in the Main Text is clearly not the tail of this peak. {Interestingly, another region of non-critical slowing down emerges in the interval `F', which is slower than the equilibration time at the previous interval `E'. The reason for further slowing away from critical pump powers is that there are is an additional condensed mode in this region, which introduces a new competition for the molecular excitation.}

\begin{figure}[]
\includegraphics[width=0.48\textwidth]{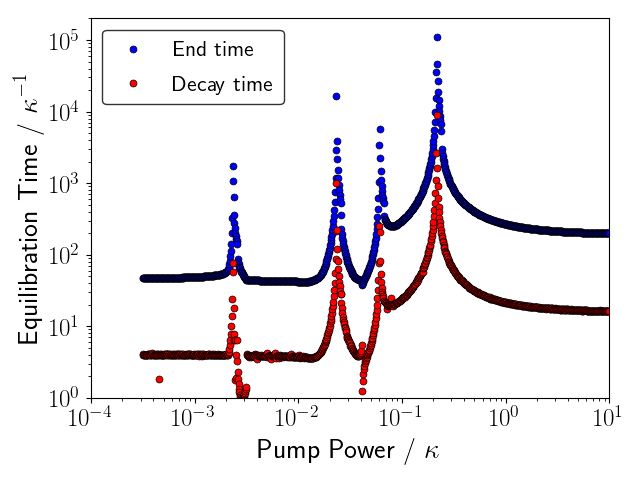}
\caption{Comparison between exponential definition of equilibration time and equilibration time based on a fixed deviation from steady state, showing the two are related by a simple scaling factor.}
\label{end_t_v_decay_t}
\end{figure}	

\section{III. Definition of equilibration time}
\label{sec3}

Equilibration time is defined in the Main Text as the time for the slowest mode to arrive at a deviation of $10^{-6}$ from its steady state value. For systems close to their steady state value, the decay towards the steady state will be exponential. In such circumstances, the exponential decay time would provide a robust definition of the equilibration time, and would be a feature of the final pump power alone, independent of the starting pump power. 

Figure~\ref{end_t_v_decay_t} shows that an exponential definition of decay time gives qualitatively the same results as those presented in Fig.~1 of the Main Text, with the two definitions related by a simple scaling factor. Later in the Main Text however, we consider equilibration times after large quenches in pump power. Under these conditions, much of the dynamics from initial state to final state is transient and non-exponential. We therefore choose a consistent definition of equilibration time which captures the key features in both cases.

\section{IV. Fitting critical power laws}
\label{sec4}

\begin{figure}[t]
\centering
\includegraphics[width=0.48\textwidth]{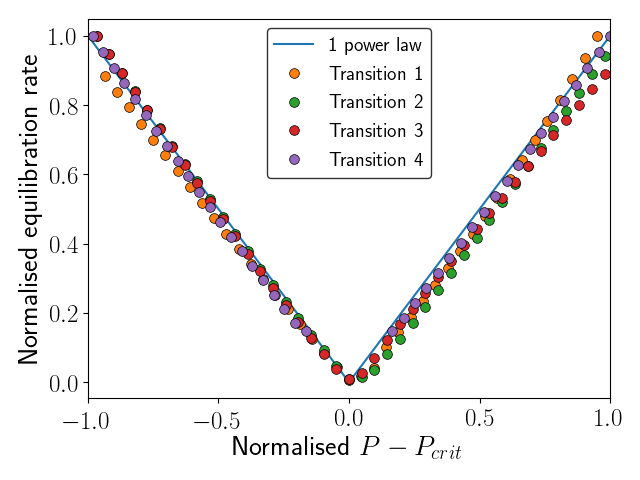}
\caption{Demonstration of the fit of the critical slowing down peaks to a critical exponent of -1. All critical peaks are rescaled in rate and $P-P_{crit}$ to make all data visible on the same plot. The four colours of data points correspond to four critical peaks in figure 1 of the Main Text. The solid line shows $|P-P_{crit}|$ as a guide to the eye.}
\label{pl_fit}
\end{figure}

Fig.~\ref{pl_fit} depicts the equilibration rate (inverse of equilibration time) as a function of the power, $P-P_{crit}$, where $P_{crit}$ is the critical power, corresponding to the four phase transitions in figure 1 of the Main Text. For critical slowing down with an exponent of -1, each phase transition should be a pair of straight lines meeting at a rate of zero for $P=P_{crit}$, as is clearly seen in figure~\ref{pl_fit}. For each critical point, because of the different power scales involved for different peaks, the values have been rescaled in $|P-P_{crit}|$ and rate to make all lines visible, and in such a way that the straight line gradient should be 1 as shown by the solid line guide to the eye in figure~\ref{pl_fit}. The coefficients of determination ($R^2$ values) for each transition, below and above threshold, are given in table~\ref{r2}.

\begin{table}[h]
\begin{tabular}{ c | c | c  }
~~~~Critical~~~~ & ~~~~Below~~~~  & ~~~~Above~~~~ \\
point & Threshold & Threshold \\ \hline
    &         &        \\
A-B & -0.9998 & 0.9985 \\
B-C & -0.9996 & 0.9992 \\
C-D & -0.9996 & 0.9999 \\
D-E & -0.9995 & 0.9998 \\
\end{tabular}
\caption{Coefficients of determination for equilibration rate as a function of $|P-P_{crit}|$ just below and just above the four critical peaks in Fig.~\ref{10_21}.}
\label{r2}
\end{table}

\bibliographystyle{prsty}
\bibliography{citations}

\end{document}